\documentstyle[12pt]{article}
\newcommand{\be}{\begin{equation}}
\newcommand{\ee}{\end{equation}}
\def\la{\mathrel{\mathpalette\fun <}}
\def\ga{\mathrel{\mathpalette\fun >}}
\def\fun#1#2{\lower3.6pt\vbox{\baselineskip0pt
\lineskip.9pt
\ialign{$\mathsurround=0pt#1\hfil##\hfil$
\crcr#2\crcr\sim\crcr}}}

\begin{document}

\begin{center}
{\bf \Large
{Relations between correlators in gauge field theory}
}
\vspace{0.5cm}
{\bf {V.I.Shevchenko\footnote{e-mail:
 shevchenko@vitep5.itep.ru}}},
{\bf {Yu.A.Simonov\footnote{e-mail:
 simonov@vitep5.itep.ru}}}

117259, Moscow,
B.Cheremushkinskaya, Russia

\end{center}

\begin{abstract}

Exact relations between gauge-invariant vacuum correlators
 in QCD are derived.
Derivatives of the correlators are expressed in terms
 of higher
orders correlators. The behaviour of the correlators
at large and small distances due to these relations 
is discussed.

\end{abstract}

\newpage

\section{}
Understanding and description of nonperturbative interactions
is one of the most important problem in the modern field 
theory.
Some progress on this way has been achieved with the method 
of
vacuum correlators (MVC) \cite{s1,s2,s3}.
Nonperturbative interactions are expressed in this method
through gauge-invariant correlators of the gluon field 
strength
operators, acting at the different space-time points.
All gauge-invariant quantities like, for example, string
tension and hadron masses may be written in terms of these
correlators \cite{s3}.
On the other hand, the latter can be computed on the lattice
 \cite{dig}, and thus play a role in the theory
as intermediate step for representing the dynamical 
information.

To have a self-consistent theoretical scheme for MVC,
it is necessary to derive the equations for correlators, 
or, at least,
some relations between different correlators.

Concerning the first, a set of exact equations 
has been proposed
\cite{a}, which
obtained by the stochastic quantization method, but, 
unfortunately
they look quite difficult to be solved and there are still 
no physical
results derived in this framework.

We choose the second way in this paper and derive simple
 relations,
yielding some information about the behaviour of 
the correlators at small and large
distances.
In the previous paper \cite{s4}
the equation for 2-point and 3-point correlators was found.
The analysis of this equation will be continued here and
the higher order
correlators - 4-point ones are also included.

We are going to study gauge-invariant Green functions (
$n$--point correlators) of the following form:
\be
\Delta^{(n)}_{\mu_1\nu_1,...,\mu_n\nu_n}=
<\>Tr(F_{\mu_1\nu_1}(z_1)\Phi(z_1,z_2)
F_{\mu_2\nu_2}(z_2)...F_{\mu_n\nu_n}(z_n)\Phi(z_n,z_1))>
\ee
here $F_{\mu\nu}(z)=F^a_{\mu\nu}(z)t^a$ -- is gluon 
field strength
and nonabelian phase factors (parallel transporters)  
$ \Phi(z,z')=P\;exp~i\int^{z'}_z A_{\mu}dx^{\mu}$ --
are crucial for gauge-invariance of the correlators.
Vacuum averages are defined as follows:
$$
<O(A)> = \int DA\> e^{-\frac{1}{4g^2}\int d^4x
Tr(F_{\mu\nu}F_{\mu\nu})}\;O(A)
$$

Any $n$--point correlator may be 
constructed using invariant tensors
$\varepsilon_{\mu\nu\sigma\rho},\> \delta_{\mu\nu}$, 
vectors
{$(z_i-z_j)_{\mu}$} and a set of scalar functions, 
depending on relative
coordinates.
For the simplest nontrivial 2--point correlator one has
two scalar functions \cite{s1}:
$$
\Delta^{(2)}_{\mu_1\nu_1,\mu_2\nu_2}=
<Tr(F_{\mu_1\nu_1}(z_1)\Phi(z_1,z_2)
F_{\mu_2\nu_2}(z_2)\Phi(z_2,z_1))>=
  $$
 $$
= \; \frac12\>\left(\frac{\partial}{\partial z_{\mu_1}}
(z_{\mu_2} \delta_{\nu_1 \nu_2} - z_{\nu_2} 
\delta_{\nu_1 \mu_2}) +
  \frac{\partial}{\partial z_{\nu_1}}
(z_{\nu_2} \delta_{\mu_1 \mu_2} - z_{\mu_2} 
\delta_{\mu_1 \nu_2})\right)\>
D_1(z_1-z_2) +
  $$
\be
+ \frac12 \> {\varepsilon}_{\mu_1\nu_1ab} 
{\varepsilon}_{\mu_2\nu_2ab}\>D(z_1 - z_2)
\ee
  Function $D(z)$ as well as $D_1(z)$  may 
be written as a sum of two parts:
  \be
  D(z)=D_{pert}(z)+D_{np}(z)
  \ee
  The first part $D_{pert}(z)\to\infty$ if $z^2\to 0$  and
  physically represents the exchange of perturbative gluons.
  Notice here, that one-gluon exchange does not get 
contribution
to the
$D_{pert}(z)$,
  the contribution to  $D_{1{pert}}(z)$  is of the 
following form:
 \be
 D_{1{pert}}(z) = 
\frac{16\alpha_s(z^2)}{3(z^2)^2}+0(\alpha^2_s).
 \ee
 At the same time for the theories with a 
nontrivial vacuum like QCD,
 there is the second part
 $D_{np}(z)$,
 which has no perturbative interpretation. 
Lattice simulations show
 \cite{dig}, that the function $D_{np}(z)$ for $|z|\ga
 0.2 fm$  falls exponentially:  
$$ D_{np}(z) \sim e^{-\frac{|z|}{T_g}} $$
 with the correlation length $T_g = 0.2 fm$,
 the behaviour of $D_{np}(z)$ for smaller $z$ is 
still unknown. At the point $z=0$
 nonperturbative components are expressed 
through the gluon condensate  \cite{s3}:
$$
\Delta^{(2)}_{\mu_1\nu_1,\mu_2\nu_2}(0)\>=
<Tr(F_{\mu_1\nu_1}\>F_{\mu_2\nu_2})>=
  $$
\be
  = \frac{1}{2}\>\varepsilon_{\mu_1\nu_1ab}
\varepsilon_{\mu_2\nu_2 ab}
  \cdot (\> D(0) + D_1(0)\>)
\ee

\section{ }

 Defined as in (2),  $\Phi(z,z')$ depends on a contour,
connecting the points $z$ and $z'$.
  It is convenient to consider a set of correlators, where 
transporters
   $\Phi$ connect neighbouring points along the 
choosen trajectories.
  We will consider correlators with the points, 
lying on a fixed
  straight line, this choice is natural for 
2--point correlator and
  we follow it for higher correlators too.

The fields under consideration are nonabelian, that's why
the derivative of the correlator with respect to $z_{\sigma}$
is not equal to the correlator of $D_{\sigma}F_{\mu\nu}(z)$
with another fields.
The difference arises from the differentiation of the 
contour and
it is crucial for the nontriviality of the picture.
Nonabelian transporters
\be
\Phi (z,z')=P~exp~i \int^{z'}_z A_{\mu} dx^{\mu}
\ee
are differentiated according to the rule \cite{co}:
\be
\frac{\partial\Phi(z,z')}{\partial z'_{\gamma}}=i
\Phi(z,z')A_{\gamma} (z') +i (z'-z)_{\rho}
\tilde{I}_{\rho\gamma}(z,z')
\ee
 where notation is used:
 $$
  \tilde{I}_{\rho\gamma}(z,z')= \int^1_0 d\alpha\;\alpha\> 
\Phi
 (z,z+\alpha(z'-z)) F_{\rho\gamma}(z+\alpha(z'-z))\cdot
 $$
 \be
 \cdot \Phi(z+\alpha(z'-z),z')
 \ee
 Analogously
\be
\frac{\partial\Phi(z,z')}{\partial z_{\gamma}}=-i
A_{\gamma} (z)\Phi(z,z') +i (z'-z)_{\rho}
{I}_{\rho\gamma}(z,z')
\ee
 $$
 {I}_{\rho\gamma}(z,z')= \int^1_0 d\alpha\cdot\alpha \Phi
 (z,z'+\alpha(z-z')) F_{\rho\gamma}(z'+\alpha(z-z'))\cdot
 $$
 \be
 \cdot \Phi(z'+\alpha(z-z'),z')
 \ee
We shall often omit the arguments of transporters below
for the sake of simplicity .
Let us consider the derivative of the 2--point
correlator (2),
using  (7) and (9) one has:

 $$
 \frac{\partial}{\partial z_{2\xi}}
\Delta^{(2)}_{\mu_1\nu_1;\mu_2\nu_2}=
 <Tr(F_{\mu_1\nu_1}(z_1) 
\Phi D_{\xi} F_{\mu_2\nu_2}(z_2) \Phi)>+
   $$
   $$
   +i (z_2-z_1)_{\sigma} \biggl( <Tr(F_{\mu_1\nu_1}(z_1)
\tilde
   I_{\sigma\xi}(z_1,z_2)F_{\mu_2\nu_2}(z_2)
\Phi(z_2,z_1))> -
   $$
   \be
   -<Tr(F_{\mu_1\nu_1}(z_1)\Phi(z_1,z_2) F_{\mu_2\nu_2}(z_2)
   I_{\sigma\xi}(z_2,z_1))>\biggr)
   \ee
The l.h.s. of this equations is:
 \be
 \varepsilon_{\mu_2\nu_2\xi\rho}
\frac{\partial}{\partial z_{2\xi}}
 \Delta^{(2)}_{\mu_1\nu_1;\mu_2\nu_2}=
 {4}\;\varepsilon_{\mu_1\nu_1
 \xi\rho}\>\frac{dD(z)}{dz^2} \; z_{\xi}
 \ee
In such a manner one obtaines a connection 
between the 2--point and 3-point
correlators:
$$
 \varepsilon_{\mu_1\nu_1\sigma\rho}
\frac{dD(z)}{dz^2}=\frac{i}{4}
 \varepsilon_{\mu_2\nu_2\xi\rho}
 \biggl(<Tr(F_{\mu_1\nu_1}(z_1)
 \tilde I_{\sigma\xi}(z_1,z_2)
 F_{\mu_2\nu_2}(z_2)\Phi(z_2,z_1))> - $$
\be
 - <Tr(F_{\mu_1\nu_1}(z_1)
 \Phi(z_1,z_2) F_{\mu_2\nu_2}(z_2)
I_{\sigma\xi}(z_2,z_1)>\>\biggr)
\ee
 We have taken into accout the Bianchi identity
 $\varepsilon_{\mu_2\nu_2\xi\rho} D_{\xi} 
F_{\mu_2\nu_2}(z)=0$
 and denoted $z_2-z_1=z$.

This relation is true for the perturbative as well as for 
nonperturbative
parts of the correlators. We will focus our attention 
on the latter,
assuming regular behaviour if their arguments tend to zero.
 The expression (13) is simplified if $z\to 0$, because in 
this case
  $ \tilde I_{\sigma\xi}(z_1,z_1)=I_{\sigma\xi}(z_1,z_1)=
 \frac{1}{2} F_{\sigma\xi}(z_1)$. Thus one obtains:
 \be
 \left. \frac{dD(z)}{dz^2}\>\right|_{z=0}=
 \frac{f^{abc}}{96}<F^a_{\mu_1\nu_1}F_{\nu_1\nu_2}^b
 F_{\nu_2\mu_1}^c>.
 \ee
 Relations (13) and (14) have been derived in \cite{s4}.
 Let us take up a question,
 what tensor structures of the 3--point correlator
contribute to the r.h.s. of (14).
There are two independent Croneker tensors in this case, 
namely:
 $$
\Delta^{(3)}_{\mu_1 \nu_1 , 
\mu_2 \nu_2, \mu_3 \nu_3 }= -i\><Tr(F_{\mu_1
\nu_1}\Phi F_{\mu_2\nu_2}
\Phi F_{\mu_3 \nu_3} \Phi)>= $$ $$
 = \frac{1}{6}\>\varepsilon_{\mu_1\nu_1 ab } 
\varepsilon_{\mu_2\nu_2 bc }
\varepsilon_{\mu_3\nu_3 ca } \cdot D_3\>+ $$ $$
+\>\frac{1}{6}\>\biggl( \varepsilon_{\mu_1\nu_1\nu_2 a } 
\varepsilon_{\mu_3\nu_3
\mu_2 a } - \varepsilon_{\mu_1\nu_1\mu_2 a } 
\varepsilon_{\mu_3\nu_3
\nu_2 a } + permut.\>\biggr)
\cdot D_2\> +
$$
\be
+ nonkroneker~ terms.
\ee
It is easy to see, that the part, proportional 
to $D_3$ does not contribute,
while $D_2$ does:
\be
f^{abc}< F^a_{\mu_1\nu_1} F^b_{\nu_1\nu_2} 
F^c_{\nu_2\mu_1}> = 48
D_2(0).
\ee
 or
 \be
 \left.\frac{dD(z)}{dz^2}\right|_{z=0} =\frac{D_2 (0)}{2}
 \ee

  Formulas (16), (17) have a simple 
interpretation in terms of the dual
Meissner effect. Introducing three-dimensional 
notations for chromoelectric
and chromomagnetic fields
$F^a_{0\alpha}=E^a_{\alpha};\; F^a_{\alpha\beta}
  =\varepsilon_{\alpha\beta\gamma}H^a_{\gamma}\;$,
   left hand side of (16) may be presented as:
\be
f^{abc}< F^a_{\mu_1\nu_1} F^b_{\nu_1\nu_2} 
F^c_{\nu_2\mu_1}> =
f^{abc}
\varepsilon_{\alpha\beta\gamma} (3<E^a_{\alpha}
E^b_{\beta} H^c_{\gamma}> + 
<H^a_{\alpha} H^b_{\beta}H^c_{\gamma}>)
\ee
Thus
a nonzero vacuum averages of triple correlators imply,
that spontaneous magnetic fluxes may be created due to
electric or magnetic lines splitting.
The existence of noncorrelated magnetic fluxes of this
 type is
equivalent to the presence of monopole currents and is 
the key point
for the possibility of confinement in such vacuum.
Notice also, that colour structure of the 3-point 
condensate
is determined by antisymmetric constants $f^{abc}$,
symmetric colour term is absent:
$$
d^{abc}< F^a_{\mu_1\nu_1} F^b_{\mu_2\nu_2} 
F^c_{\mu_3\nu_3}> = 0
$$
or
$$
< F^a_{\mu_1\nu_1} F^b_{\mu_2\nu_2} F^c_{\mu_3\nu_3}> = 
\frac{f^{abc}}{6}\>
\Delta^{(3)}_{\mu_1\nu_1 , \mu_2\nu_2 , \mu_3\nu_3}(0)
$$

The equation (13) will be examined here in more details.
One can prove, that it is equivalent to :
\be
z^2\>\frac{dD(z)}{dz^2}\>=\>\frac{1}{2}\;
\int_0^z tdt\> \biggl(\>D_2(z-t,
 t)\> + \> D_2(-t, t-z)\biggr) \ee
Let us make an assumption, that function $D$
and $D_2$
may be parametrized,
taking into account their cyclic symmetry, as:
$$ D(z)\>
=\>D(0)\> h(z)h(-z);\;\; D_2(z-t, t)\> =\>D_2(0)\>
h_2(z-t)h_2(t)h_2(-z) $$ Then, it is easy to check, 
that the solutions of
the following equation satisfy also (19):
\be
h(-z)\>\frac{dh(z)}{dz}\>=\>- \frac{a^2}{4}\>h_2(-z)\;
\int_0^z dt\>
h_2(t)\cdot h_2(z-t)
\ee
where $a^2 = - \frac{2\>D_2(0)}{D(0)}$.
Naively one can think, that $h(z) = h_2(z)$ for all $z$.
Equation (20) may be solved in this case and the regular 
at $z=0$
solution has the form:
$$
h(z)\>=\> 2\;\frac{J_1(az)}{az}
$$
where $J_1(az)$ - Bessel function of the first type.
This behaviour contradicts the lattice measurments 
\cite{dig},
it means, that the relation  $h(z)\>\equiv\>h_2(z)$
does not hold in the real gluodynamic vacuum.
Still this solution leads to the definite value of 
the string
tension, $\sigma$:
$$
\sigma = \int\limits_0^{\infty} d^2 z\> D(z) = 
\frac{2\pi \left(D(0)\right)^2}{D_2(0)}
$$
The important point here is the negative sign 
of the fraction
$\frac{D_2(0)}{D(0)}$, i.e. $a$ should be real, overwise
correlation functions would unphysically increase 
with a distance.
Indeed, there are several grounds to think, 
(see \cite{s4} and ref. in \cite{dig})
that the 3-point condensate is negative in the real world.

Let us consider a more realistic case $h(z) \neq h_2(z)$.
Lattice results allow to look for the solution in the form:
$$
D(z) = D(0)\> e^{-\frac{|z|}{T_g}}\>
\sum_{n=0}^{\infty}
\frac{d_n}{n!}{\left(\frac{z}{T_g}\right)}^n
$$
\be
D_2(x, z-x) = D_2(0)\>e^{-\frac{|x|}{T_g}}\>
\sum_{m=0}^{\infty}
\frac{b_m}{m!}{\left(\frac{x}{T_g}\right)}^m
\>\cdot
e^{-\frac{|z-x|}{T_g}} \> 
\sum_{m=0}^{\infty}
\frac{b_m}{m!}{\left(\frac{z-x}{T_g}\right)}^m
\ee
The substitution of (21) in Eq. (19) gives a 
relation between $d_n$ and $b_m$:
\be
d_{n+2} =  d_{n+1} - \frac{(a\>T_g)^2}{2}
\;\sum_{m=0}^{n} b_m\> b_{n-m}
\ee
with the obvious conditions $d_0 = b_0 = d_1 = 1 $.
There are infinitely many sets of 
coefficients $b_m , d_m$, which solve
the above equation. What solution is 
realized in the real world is a question
for future investigations.

\section{ }

It remains to be seen whether the above picture 
is compatible with the
Gaussian model for QCD vacuum
\cite{s1}, \cite{nacht}.
In the latter all correlators of the odd orders, 
including
 $\Delta^{(3)}$ are assumed to be zero, 
while the even correlators are factorized
in a product of the 2--point ones.
Among main arguments in favour of it is the fact, that
 potential of confinement for higher representations 
of a gauge group
is proportional to the quadratic Casimir operator 
in the Gaussian vacuum.
This fact is confirmed with a good accuracy by 
lattice computations (see
discussion in \cite{ufn}). The Gaussian model for QCD 
vacuum was introduced in
\cite{s1} and intensively used in \cite{dokr} 
for the high energy scattering
amplitudes.
Eq. (19) shows clearly, that such picture is not 
selfconsistent
for any nontrivial function
$D(z)$.
Thus one should introduce an extended Gaussian ensemble 
and postulate
factorization in a product of the 2--point and 3--point 
correlators.
If 3--point  correlator is zero, this definition 
coincides with the
ordinary Gaussian ensemble,  because all odd 
order correlators are also zero, in
the opposite case a new stochastic ensemble 
arises, it is natural to
call it the minimally extended Gaussian ensemble.
At the same time even in the ordinary Gaussian 
vacuum function
 $D(z)$ may be a nontrivial one.
We shall deal in this case with the second derivative 
of the 2--point
correlator at $z=0$:
 $$ \left.\frac{\partial^2
\Delta^{(2)}}{\partial z_{2\xi } \partial
z_{1\rho}}
\varepsilon_{\mu_1\nu_1\rho\eta}
\varepsilon_{\mu_2\nu_2\xi\gamma}\right|_{z_1\to z_2} =
\varepsilon_{\mu_1\nu_1\rho\eta}
\varepsilon_{\mu_2\nu_2\xi\gamma}\times
$$
$$
\times \biggl[-\frac{i}{2}\><Tr(F_{\mu_1\nu_1}[F_{\rho\xi}
F_{\mu_2\nu_2}]) > +\> \frac{i\>z_{\sigma}}{6}\> 
<Tr(F_{\mu_1\nu_1}[D_{\rho}
F_{\sigma\xi} F_{\mu_2\nu_2}]) > -
$$
$$
-\frac{z_{\sigma}z_{\phi}}{24} \biggl(
6 \> <Tr(F_{\mu_1\nu_1}
F_{\phi\rho} [F_{\sigma\xi} F_{\mu_2\nu_2}]) >
+ 6 \> <Tr(F_{\mu_1\nu_1}
[F_{\mu_2\nu_2} F_{\sigma\xi}] F_{\phi\rho})> +
$$
\be
 + \> <Tr(F_{\mu_1\nu_1}
 [F_{\sigma\xi} F_{\phi\rho}] F_{\mu_2\nu_2}) >
+ \> <Tr(F_{\mu_1\nu_1}
F_{\mu_2\nu_2} [F_{\phi\rho} F_{\sigma\xi}]) > 
\biggr)\biggr]
\ee
Taking into account the following:
\be
\frac{\partial^2\Delta^{(2)}(z)}{\partial z_{2\xi } 
\partial
z_{1\rho}}
\varepsilon_{\mu_1\nu_1\rho\eta}
\varepsilon_{\mu_2\nu_2\xi\gamma}=
16\;
\frac{d^2D(z)}{(dz^2)^2}\>
(z_{\gamma}z_{\eta}-\delta_{\gamma\eta}z^2)\>
- \> 24 \;\frac{dD}{dz^2}\>\delta_{\gamma\eta}
\ee
one has:
$$
-48 D^{\prime\prime}|_{0} = \lim_{z\to
0}\>\biggl[\;\frac{i z_{\sigma}}{6 z^2}
\varepsilon_{\mu_1\nu_1\rho\eta}
\varepsilon_{\mu_2\nu_2\xi\eta}
 <Tr(F_{\mu_1\nu_1} [ D_{\rho} F_{\sigma\xi}, 
F_{\mu_2\nu_2}])>-
$$
$$
-\frac{z_{\sigma}z_{\phi}}{24 z^2} \biggl(
6 \> <Tr(F_{\mu_1\nu_1}
F_{\phi\rho} [F_{\sigma\xi} F_{\mu_2\nu_2}]) >
+ 6 \> <Tr(F_{\mu_1\nu_1}
[F_{\mu_2\nu_2} F_{\sigma\xi}] F_{\phi\rho})> +
$$
\be
 + \> <Tr(F_{\mu_1\nu_1}
 [F_{\sigma\xi} F_{\phi\rho}] F_{\mu_2\nu_2}) >
+ \> <Tr(F_{\mu_1\nu_1}
F_{\mu_2\nu_2} 
[F_{\phi\rho} F_{\sigma\xi}]) > \biggr)\>\biggr]
\ee

The first and the second terms in the r.h.s. 
of (23) are absent
in the Gaussian vacuum, 4--point condensate has to 
be introduced for
the analysis of the others:
$$
\Delta^{(4)}=<Tr(F_{\mu_1\nu_1}\Phi F_{\mu_2\nu_2}\Phi
F_{\mu_3\nu_3}\Phi F_{\mu_4\nu_4}\Phi)>= D_{4}^{connect} +$$
$$
+ (\alpha_0\cdot\;
\varepsilon_{\mu_1\nu_1 ab}\> \varepsilon_{\mu_2\nu_2
bc}\; 
\varepsilon_{\mu_3\nu_3 cd}\>
\varepsilon_{\mu_4\nu_4 dc} + \alpha_1
\cdot\;\varepsilon_{\mu_1\nu_1ab}\>
\varepsilon_{\mu_3\nu_3ba}\;
\varepsilon_{\mu_2\nu_2cd}\>\varepsilon_{\mu_4\nu_4 dc} +
$$
\be
+ \alpha_2\cdot\;
\varepsilon_{\mu_1\mu_2ab}\>\varepsilon_{\mu_4\nu_4ba}\;
\varepsilon_{\mu_2\nu_2cd}\>\varepsilon_{\mu_3\nu_3dc})
\cdot
D_{4}^{Gauss} + nonkron.\; terms
\ee
here the connected 4--point function
 $D_{4}^{connect}$ should also be 
zero in the Gaussian vacuum.

Using (5):
\be
<F^a_{\mu_1\nu_1} F^b_{\mu_2\nu_2}>= 
\frac{\delta^{ab}}{8}
\varepsilon_{\mu_1\nu_1\rho\sigma}
\varepsilon_{\mu_2\nu_2\rho\sigma} (D(0)+D_1(0))
\ee
it is easy to get:
\be
\alpha_0=\alpha_2=\frac{16}{3};~\alpha_1=
-\frac{2}{3};~ D_{4}^{Gauss}=
\left(\frac{D(0)+D_1(0)}{8}\right)^2
\ee
Straightforward calculations lead to the result:
\be
\left.\frac{d^2D(z)}{(dz^2)^2}\right|_{z=0}=
\frac{7}{2}\left(\frac{D(0)+D_1(0)}{4}\right)^2
\ee

 We see, that even for the  Gaussian ensemble 
the presence of the nonperturbative
condensate
\be
D(0)+D_1(0)\neq 0
\ee
automatically provides the dynamical dependence 
on $z$ for function
$D(z)$.
\section{ }

The relation (29)
demonstrates the existence of a scale, 
which determines the behaviour
of
$\Delta^{(2)}(z)$ at small distances and 
it may be quite different from
the asymptotic
correlation length $T_g$.
Indeed, taking from \cite{dig} $D_1(z)
\approx \frac13 D(z)$ and defining 
$D(0) + D_1(0)$ from the standard
gluon condensate \cite{nacht} , 
we get $$ \frac{D''(0)}{D(0)} \approx 0.012\;
Gev^{4} $$ whence it follows 
that (parametrizing function $D(z)$
 near zero as a gaussian)
 $$
 D(z) = D(0)e^{- z^2/r_0^2}
$$
and $r_0 \approx 0.6 fm$.
This value of $r_0$ is larger than $T_g \approx
0.2 fm$,  characterizing the 
exponential asymptotic decreasing of the
correlator for $z > 0.2 fm$:
$$ D(z) =
D(0)e^{- |z|/T_g}
$$
if $ z > 0.2 $ ä¬.

This situation is possible in the so called string limit:
$$
T_g^2(D(0)+D_1(0))\sim \sigma= const
$$
with $T_g \to 0$.
It is seen from (29), that also:
%\be
%T_g^4(D(0)+D_1(0)) \to 0.
%\ee
\be
\left.\frac{T_g^4}{D(0) + D_1(0)}\>
\frac{d^2D(z)}{(dz^2)^2}\right|_{z=0}
\to 0
\ee
Substituting in (31) values for 
$D(0) + D_1(0)$ and $T_g$ from \cite{dig},
we get on the l.h.s. $0.04 
\sim {\left(\frac{T_g}{r_0}\right)}^4,$ that
makes string limit in the 
nonperturbative QCD credible.

Let us make another 
important notice about the Gaussian approximation
for the gluodynamic vacuum.
The Taylore expansion 
for even function $D(z)$ is as follows:
$$
D(z) = \sum_{n=0}^{\infty} 
\left.\frac{d^n D(z)}{(dz^2)^n}\right|_{z=0}
\frac{z^{2n}}{n!}
$$
The coefficients of the above 
expansion are expressed through the
$(2+n)$-th and less orders condensates. 
We have seen that for $n=1$ and
$n=2$ the derivatives at $z=0$ 
are zero and positive respectively for the
Gaussian ensemble. It may be interesting to 
check, are there negative
coefficients in this Taylor expansion for 
the Gaussian vacuum or not. The
latter would be obviously incompatible 
with the observed decreasing of the
function $D(z)$ with $z$ and 
would be a rigorous proof of nongaussian
nature of gluodynamic vacuum.

The behaviour of the 
function $D(z)$ for $z\la 0.2 fm$ is unknown
at the moment.
Its determination and higher 
correlators measurments would take a chance
to check the relations
(13), (14), (22) and make a 
conclusion, what type of vacuum is realized
in gluodynamics and 
whether one can approximate this vacuum as
the Gaussian or the minimally 
extended Gaussian ensemble of the stochastic fields.
We have said nothing  about perturbative 
parts of correlators
and their connection with the 
nonperturbative ones. This important
question will be considered in its own right.

\begin{flushleft}
{\bf Acknowledgements}
\end{flushleft}
This work was supported in part by 
the RFFI grants No 96-02-19184 and
No 96-02-04808.

\end{document}